\documentclass[conference]{IEEEtran}
\IEEEoverridecommandlockouts
\usepackage{cite}
\usepackage{amsmath,amssymb,amsfonts}
\usepackage{algorithmic}
\usepackage{graphicx}
\usepackage{textcomp}
\usepackage{xcolor}
\usepackage{booktabs}
\usepackage{hyperref}
\def\BibTeX{{\rm B\kern-.05em{\sc i\kern-.025em b}\kern-.08em
    T\kern-.1667em\lower.7ex\hbox{E}\kern-.125emX}}
\begin{document}

\title{Classification of Emerging Neural Activity from Planning to Grasp Execution using a Novel EEG-Based BCI Platform\\
\thanks{Supported by RI-INBRE (NIH P20GM103430) \& NSF award ID 2245558.}
}

\author{
    \IEEEauthorblockN{Anna Cetera, Ali Rabiee, Sima Ghafoori, Reza Abiri}
    \IEEEauthorblockA{
        Department of Electrical, Computer and Biomedical Engineering, University of Rhode Island, Kingston, RI, USA\\
        Emails: \{annacetera, ali.rabiee, sima.ghafoori, reza\_abiri\}@uri.edu
    }
}


\maketitle


\begin{abstract}
There have been different reports of developing Brain-Computer Interface (BCI) platforms to investigate the noninvasive electroencephalography (EEG) signals associated with plan-to-grasp tasks in humans. However,
these reports were unable to clearly show evidence of emerging neural activity from the planning (observation) phase - dominated by the vision cortices - to grasp execution - dominated by the motor cortices. In this study, we developed a novel vision-based-grasping BCI platform that distinguishes different grip types (power and precision) through the phases of plan-to-grasp tasks using EEG signals. Using our platform and extracting features from Filter Bank Common Spatial Patterns (FBCSP), we show that frequency-band specific EEG contains discriminative spatial patterns present in both the observation and movement phases. Support Vector Machine (SVM) classification (power vs precision) yielded high accuracy percentages of 74\% and 68\% for the observation and movement phases in the alpha band, respectively.
\end{abstract}



\section{Introduction}
The ability to independently navigate and interact with the surrounding environment through reaching and grasping is fundamental to an individual’s autonomy and quality of life. Neural correlates associated with natural reach-and-grasp actions can be decoded and identified through invasive electrocorticography (ECoG) \cite{VargasIrwin2015}, offering insights into the emergence of neural activity before and during movement onset.

While current ECoG-based Brain-Computer Interface (BCI) systems have been explored for assisting individuals with grasp disabilities, existing control methods often lack a natural and intuitive feeling of control \cite{ajiboye2017restoration, raichle2008prosthesis}. Common laboratory experimental set-ups attempt to employ a naturalistic reach-to-grasp set-up by presenting multiple objects to the participant simultaneously \cite{schwarz2017decoding, schwarz2020analyzing}. In these setups, the lack of object isolation may introduce bias to the data, creating difficulties in understanding the relationship between the visual and motor cortices linked to individual objects and their specific grip types.

To eliminate the possibility of data bias, we employed a novel EEG vision-based-grasping platform that distinguishes the neural activity between the observation (planning) phase and movement (grasp execution) phase through vision control and object isolation. Gaining control over the timing at which the participants observe the object presented before them enables us to analyze the emergence of neural activity across different phases of plan-to-grasp tasks. Meanwhile, isolating the object presented before them allows for the exploration of distinct neural patterns associated with the object and its specific grip type (power vs precision). 

In this study, we aim to achieve two primary goals. Firstly, to obtain results that are consistent with the frequency-band-specific neural spatial patterns reported in ECoG studies to validate our platform, particularly within the alpha band due to its strong association with motor planning \cite{mcfarland2000mu, pfurtscheller1997existence, li2018combining}. Secondly, to introduce the possibility of additional frequency-band-specific features that may be involved with motor planning during the observation phase without explicit motor imagery instructions \cite{xu2021decoding, wang2004classifying}. By implementing FBCSP, we examine the frequency-specific-spatial patterns during plan-to-grasp tasks while classifying between grasp types using SVM during the observation and movement phases within specific frequency ranges.

Investigating the emerging neural activity from the planning (observation) phase - dominated by the vision cortices - to grasp execution - dominated by the motor cortices within specific frequency bands may offer new insight for a naturalistic control strategy for noninvasive BCI systems and improve the quality of life for those with impaired hand dexterity.


\section{Materials and Methods}
\subsection{Data Collection Platform }
In this study, two distinct objects were selected to execute two specific reach-to-grasp actions most used in daily life: the precision and power grasp types. The objects were (i) a pen for precision grasp execution and (ii) a water bottle for power grasp execution (Figure 2b). The object was presented to the participant on a sectioned, motorized turntable while wearing a pair of developed "smart glasses" capable of transparency alternation (Figure 1a). To eliminate the possibility of data bias, no object was presented between object A and object B, the selection of the objects was randomized, and the smart glasses transitioned from a transparent to an opaque state during the rotation of the turntable. \\
\\

\textit{Hardware}

The presentation of object A, object B, and no object required a novel, PC-controlled, motorized turntable divided into 3 sections. Each object  was individually placed in one of the three sections (Figure 1b). Two Arduino Uno microcontrollers were programmed to control the motor driver (TB6600 4A 9-42V Stepper motor driver), the motor (Bipolar 1.7A Nema 17 Stepper Motor), the audio cue, and the “smart glasses.” The developed “smart glasses” enabled/disabled object visibility due to the transparent/opaque capabilities of PDLC electronic smart film.  

\textit{Software} 

The developed software ensured synchrony between the EEG data acquisition and the hardware components for accurate event logging. PySerial was utilized to send numerical commands from Python to Arduino IDE to control audio cues and change the states of the motorized turntable/smart glasses. A simple, real-time graphical user interface (tkinter) was additionally developed to run in synchrony with the platform at the time of data collection for the researcher (Figure 1c).

\begin{figure}[ht!] 
\centering
\includegraphics[width=3.5in]{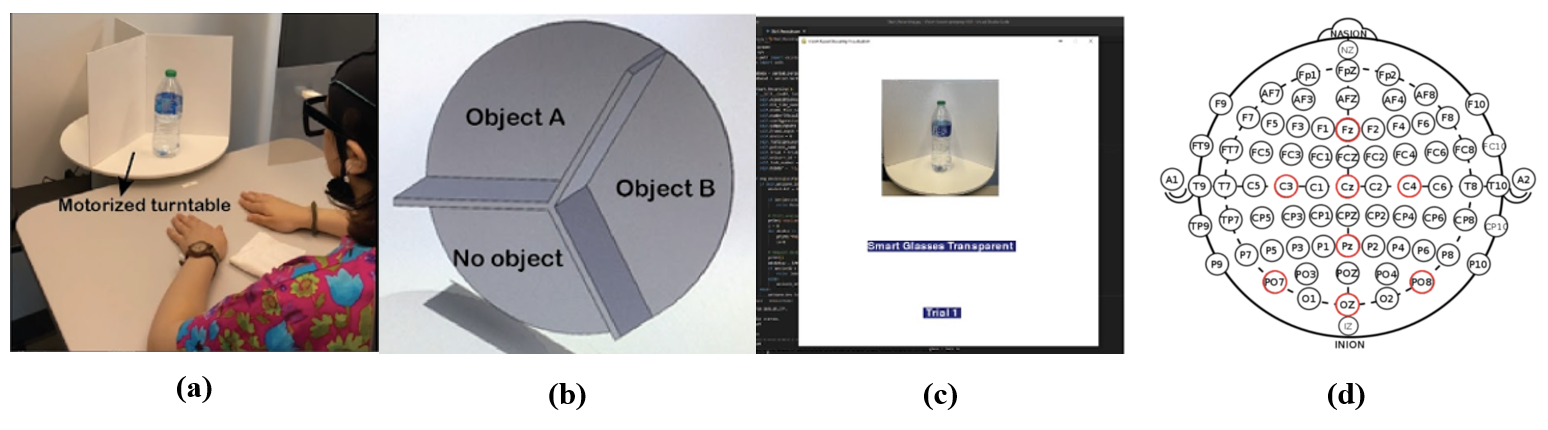}
\caption{Experimental setup and EEG electrode placement. (a) Participant wearing EEG headset and "smart glasses" seated in front of a motorized turntable with an object to perform a reach-to-grasp task. (b) 3D-designed motorized turntable with object A, object B, and no object sections. (c) Synchronized graphical user interface during data collection. (d) Standard EEG 10-20 system electrode placement with highlighted electrodes (red) used for analysis.}
\label{experimental_paradigm}
\end{figure}

\subsection{Data Collection}
\textit{Participants}

Data was collected from five human subjects (3 female, 2 male) aged between 20 and 35. Each participant was right-hand dominant with no known motor deficits or neurological impairments. Each participant attended a single data collection session (approximately 75 minutes).

\medskip
\textit{Data Acquisition}

EEG signals were acquired using the 8-channel Unicorn Hybrid Black headset with a wet electrode setup (manufactured by g.tec) at a sampling rate of 250 Hz. Electrodes were placed in the following positions according to the international 10/20 system: Fz, C3, Cz, C4, Pz, PO7, Oz, PO8 (Figure 1d). Reference and ground electrodes were placed on the left and right mastoid respectively. Quality testing of EEG signals was performed before each data recording session within the Unicorn Suite Hybrid Black software environment (g.tec).

\medskip
\textit{Protocol Design} 

In our study, the participant was instructed to perform a reach-to-grasp task with enabled vision capabilities and motor movement execution. The participant was positioned 30cm from the center of the object with their palms facing downwards. The structure of the data collection protocol is centered on an audio-cue-based paradigm, where each data recording session included 5 blocks, each containing 10 trials, resulting in a total of 50 trials for each object presented (Figure 2). 

Before the audio cue, the participant was instructed to observe the object for two seconds. Following the audio cue, the participant performed a naturalistic reach-to-grasp task toward the object presented by the motorized turntable for four seconds. If no object was presented, the participant was instructed to execute no movement.

\begin{figure}[ht!] 
\centering
\includegraphics[width=3.5in]{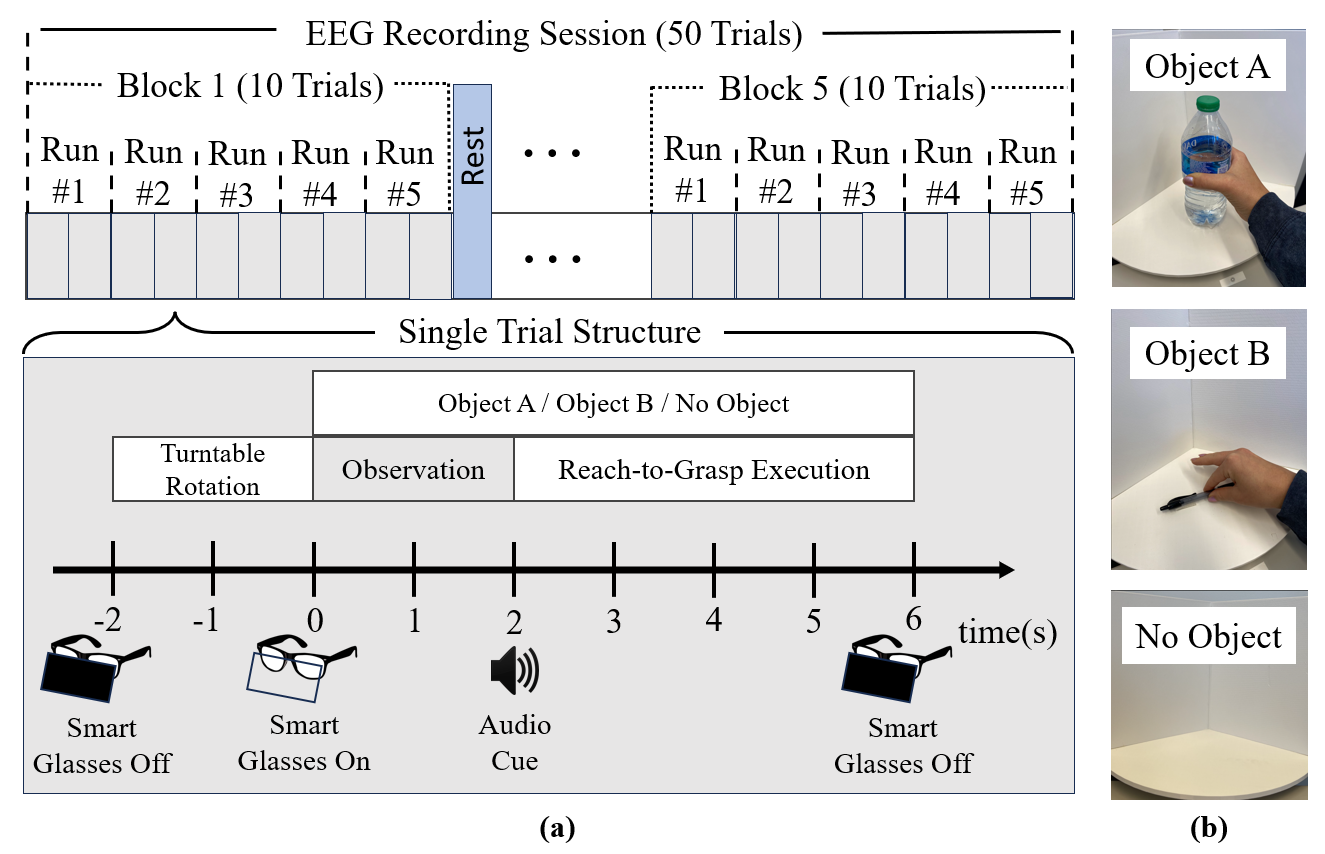}
\caption{Structure of data collection session and singular EEG trials. (a) One collection session for each subject involves multiple blocks, rest periods, and the sequence of events within a single trial including the periods of turntable rotation, object observation, reach-to-grasp execution, and auditory cues. (b) Presented objects on the motorized turntable during the experiment: object A (a water bottle), object B (a pen), and the 'no object' condition.}
\label{experimental_paradigm}
\end{figure}

\subsection{Data Processing and Feature Extraction}
The acquired EEG data was initially filtered with a 60 Hz notch filter to suppress power line noise. To eliminate low-frequency drift, a zero-phase, 4th-order Butterworth bandpass filter with cutoff frequencies of 0.5 and 40 Hz was applied. To implement the Filter Bank Common Spatial Pattern algorithm, subject-specific and object-specific single trial EEG were decomposed into the following filter banks: (delta: 0-4Hz, theta: 4-8Hz, alpha: 8-13Hz, beta: 13-30Hz, gamma: 30-40Hz), using zero-phase 4th-order Butterworth bandpass filters. 

Two windows of interest were extracted for binary classification: two seconds before the audio cue (observation phase) and two seconds after the audio cue (movement phase). The filtered, epoched data was used to calculate the CSP projection matrix for each object within each filter bank. The first and last two filter components were extracted to spatially filter the raw, single-trial EEG data. The feature set included the logarithms of normalized variances from the most and least discriminative spatial components, which respectively maximize and minimize variances for each class. In each filtered trial, four CSP features were derived and implemented for binary classification.

\subsection{Classification}
The FBCSP features were utilized for binary classification to distinguish between the power grasp and precision grasp during the observation and movement phase within different frequency bands. The Support Vector Machine (SVM) algorithm was used to perform binary classification. Table I displays subject-based binary classification results (percent accuracy) within each frequency band during both the observation and movement phases of the experiment. 

\section{Results}

\subsection{Common Spatial Pattern}

The CSPs extracted from the alpha band during the observation and movement phases are shown in Figure 3. During the observation phase, CSP \#1 displays an increase of alpha power within the temporal and occipital regions of the brain (indicated by red) while being surrounded by lower alpha power (indicated by blue) in the frontal regions of the brain. CSP \#2 features lateralized activity as lower and higher alpha band power are localized over the left and right hemisphere regions respectively. Both CSP \#3 and CSP \#4 display reduced alpha activity in the occipital region, while CSP \#1 and CSP \#2 display increased alpha activity in the same location. 

The first common spatial pattern during the movement phase (CSP \#1) exhibits localized alpha activity in the central region of the brain. CSP \#3 and CSP \#4 display an increase of alpha activity in the occipital region, however, opposite alpha activities occur across the motor cortex region between the two patterns.

\begin{figure}[ht!] 
\centering
\includegraphics[width=3.5in]{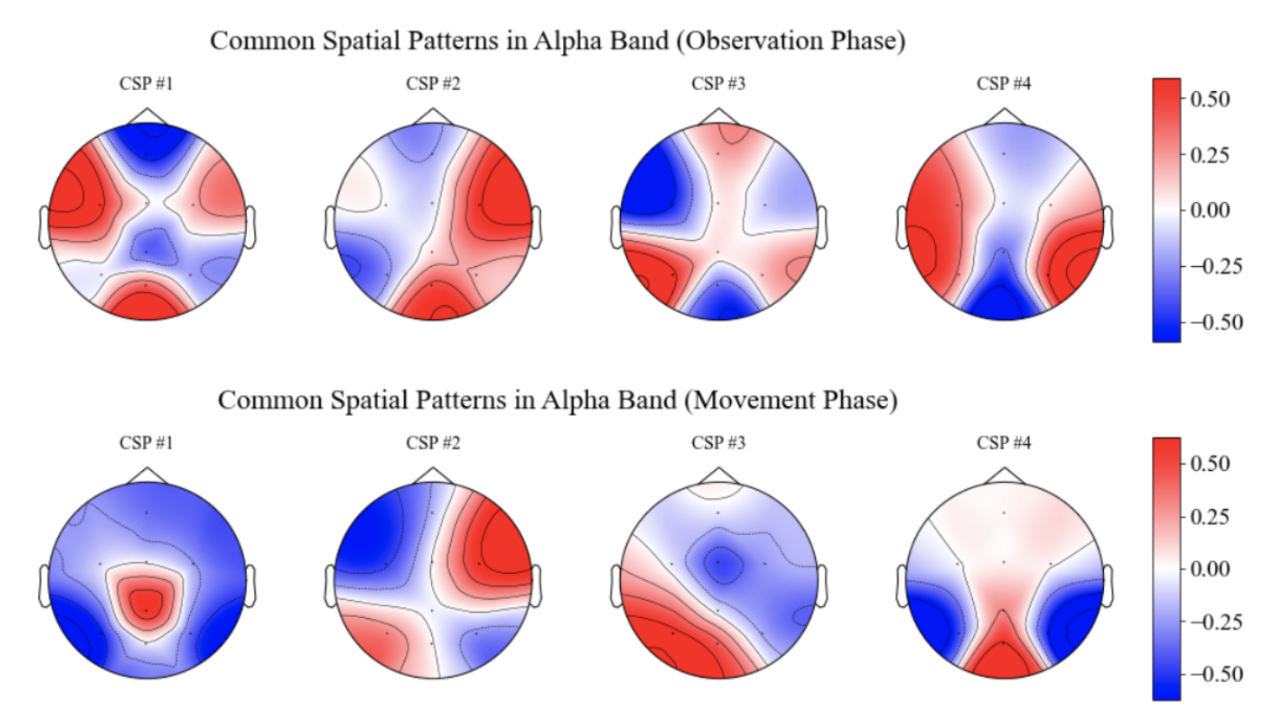}
\caption{Topographical plots of extracted common spatial patterns (CSPs) in the alpha band during different task phases for subject 3. The upper panel illustrates CSPs \#1 to \#4 during the observation phase. The lower panel displays the corresponding CSPs during the movement phase. Each map is scaled from -0.50 to +0.50, reflecting changes in amplitude within the alpha band}
\label{experimental_paradigm}
\end{figure}

\subsection{Classification}

The classification results in Table I display that analysis of the mean accuracies for binary classification between power grasp and precision grasp during both the observation and movement phases yields insightful trends, particularly in the alpha, delta, and gamma frequency bands. Across all subjects, the features within the alpha band contributed to the highest mean accuracy 74\% in the observation phase and 67\% in the movement phase. Furthermore, mean accuracies in the delta and gamma bands demonstrate an increase during the observation phase compared to the movement phase, with delta band accuracies rising from 55\% to 68\% and gamma from 63\% to 65\%. 

The box-and-whisker plots in Figure 4 depict the distribution of SVM classification accuracies across different EEG filter bank frequency bands during the observation and movement phases. Upon initial observation, there is an increase in overall percent accuracy across all subjects during the observation phase (except for the gamma band), contributing to the separation between the two phases. The alpha band exhibits the highest median accuracy during both phases across all frequency bands, suggesting that features derived from the alpha band are the most discriminative for classifying grip types. The median accuracies in the alpha band surpass those in the delta, theta, beta, and gamma bands, which indicates it as a consistent feature with less variability in the classification performance in distinguishing between the power and precision grasps across different trials or subjects.

\begin{table*}[ht!]
\centering  
\caption{Subject-Based Grip Type Classification during Observation and Movement Phase}
\label{tab:observation_movement}
\resizebox{\textwidth}{!}{%
\begin{tabular}{lcccccccccccc}
\toprule
 & \multicolumn{5}{c}{Observation Phase (\%)} & \multicolumn{5}{c}{Movement Phase (\%)} \\
\cmidrule(r){2-6} \cmidrule(l){7-11}
Subjects & Delta & Theta & Alpha & Beta & Gamma & Delta & Theta & Alpha & Beta & Gamma \\
\midrule
s1 & 45 & 55 & 80 & 50 & 60 & 45 & 50 & 65 & 45 & 70 \\
s2 & 75 & 60 & 70 & 60 & 50 & 65 & 40 & 60 & 50 & 45 \\
s3 & 80 & 65 & 70 & 75 & 65 & 60 & 65 & 80 & 80 & 75 \\
s4 & 80 & 60 & 85 & 60 & 70 & 65 & 75 & 75 & 65 & 60 \\
s5 & 60 & 65 & 65 & 75 & 55 & 40 & 65 & 55 & 55 & 65 \\
\midrule
Mean & 68 & 61 & \textbf{74} & 64 & 65 & 55 & 59 & \textbf{67} & 59 & 63 \\
\bottomrule
\end{tabular}
}
\end{table*}

\begin{figure}[ht!] 
\centering
\includegraphics[width=3.5in]{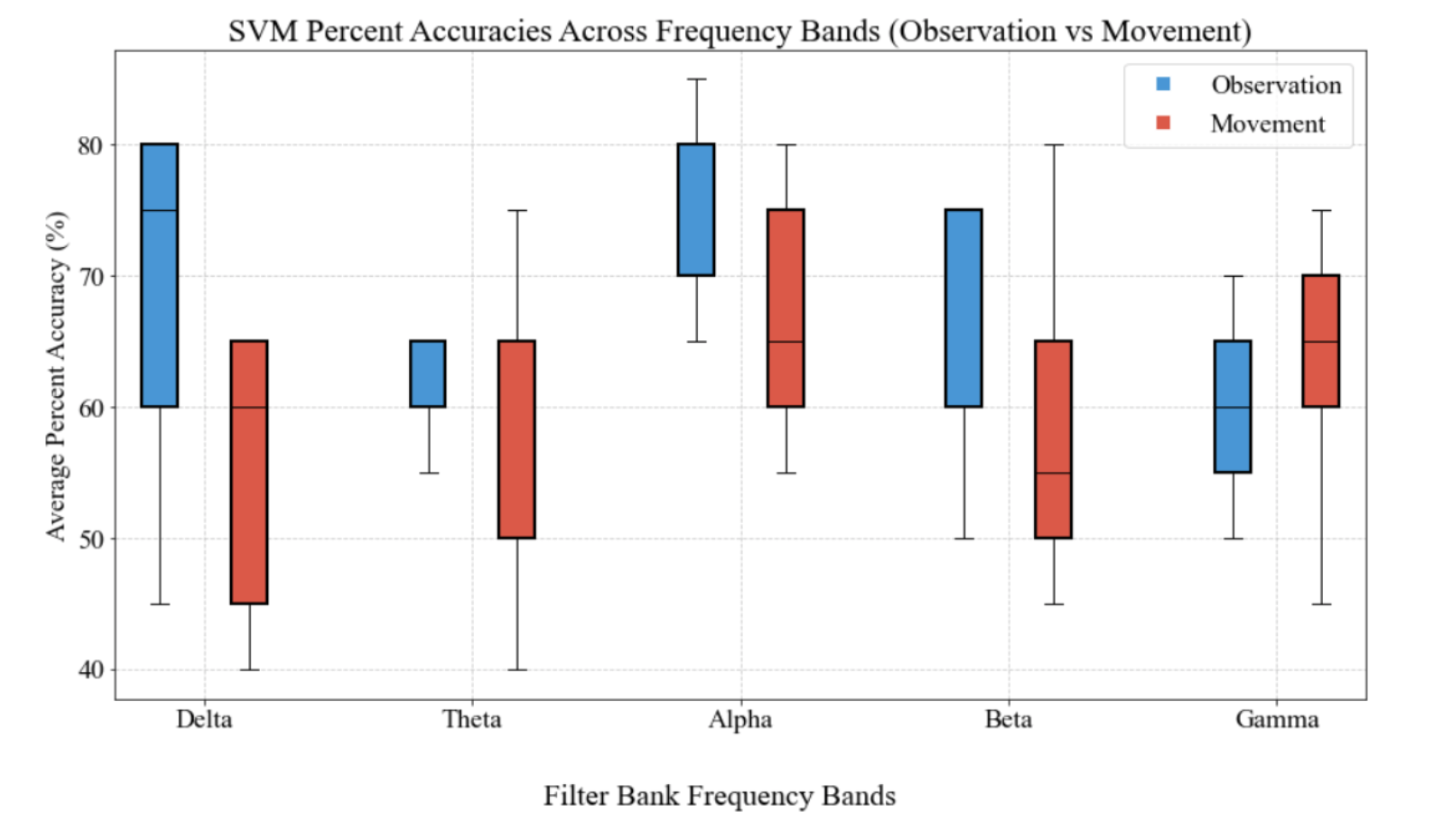}
\caption{Box-and-whisker plots illustrating the distribution of SVM classification accuracies across five frequency bands during the observation and movement phases. Blue boxes represent accuracies during the observation phase, while red boxes correspond to the movement phase.}
\label{experimental_paradigm}
\end{figure}


\section{Discussion}

Our novel platform's ability to separate phases during plan-to-grasp tasks and to isolate objects allows us to observe distinct trends in neural activity during the observation and movement phases of reach-to-grasp tasks respective to the specific object presented. The derived CSP topographies, specifically in the alpha band during the observation phase, could have significant implications for understanding motor planning. Alpha activity is known to be associated with motor planning \cite{pfurtscheller1997existence}, and the trends observed in our study are consistent with this connection since the detected ERD during the observation phase is evident in our results. ERD is present within the occipital regions of the brain (responsible for processing visual stimuli) where decreased levels of alpha activity are present within the third and fourth CSPs during the observation phase.

The SVM classification accuracies derived from the FBCSP features, especially within the alpha band, demonstrated the highest percent accuracy across all subjects during the observation phase is also consistent with previous literature in that this particular frequency band plays a pivotal role in discriminating between the 2 grasps types. These findings validate our novel platform, while also introducing the possibility of exploring delta and gamma frequency bands that may be involved in motor planning without explicit motor imagery instructions due to their high accuracy percentages based on our findings. Although these accuracy percentages are less significant than those observed in the alpha band, they could be indicative of the delta band's association with integrative sensory processing and the gamma band's link to higher-level cognitive functions. The increased accuracies in these bands during the movement phase may reflect the increased demand on sensorimotor integration and the heightened cognitive engagement required for executing the motor task.


\section{Conclusion}

The ability to control vision and separate the observation phase from the movement phase during reach-to-grasp tasks with our novel platform enabled us to uncover neural mechanisms and activity associated with object-specific visually guided tasks using noninvasive EEG. Isolating the object presented introduced the possibility of novel findings regarding how delta and gamma frequency activity might play a role in motor planning for grasp classification. Without explicit motor imagery instructions, we were able to detect significant neural patterns during the observation phase. 

The outcomes gained from performing FBCSP/SVM on the EEG data collected from our novel platform for grip type classification open the possibility of exploration of frequency band-specific neural activity that is solely associated with the singular object presented to the participant. This platform facilitates the investigation for understanding the interplay between different brain regions at various frequency bands during motor planning and execution, particularly in the context of complex tasks like power and precision grasps. This novel vision-based-grasping platform presents a new direction for non-invasive BCI systems by exploring the emergence of neural activity during motor planning and visually guided tasks.




\bibliographystyle{IEEEtran}
\bibliography{main}

\end{document}